\documentclass[10pt,conference]{IEEEtran}
\IEEEoverridecommandlockouts
\usepackage{cite}
\usepackage{amsmath,amssymb,amsfonts}
\usepackage{algorithmicx}
\usepackage{algpseudocode}
\usepackage{graphicx}
\usepackage{textcomp}
\usepackage{colortbl}
\usepackage{algorithm}
\usepackage{framed,enumitem}
\usepackage{ragged2e}
\usepackage{booktabs}
\usepackage{multirow}
\usepackage{newtxtext,newtxmath}
\usepackage[table]{xcolor}
\usepackage{tabularray}
\UseTblrLibrary{booktabs}
\usepackage{array}

\usepackage[most]{tcolorbox}
\tcbuselibrary{skins}

\definecolor{memtitlebg}{HTML}{E8F4F1}
\definecolor{memframe}{HTML}{7DA8A1}
\definecolor{ctxgreen}{HTML}{2E8B57}
\definecolor{feedred}{HTML}{C44E52}
\usepackage{hyperref}
\tcbset{
  memorycard/.style={
    enhanced,
    breakable,
    colback=white,
    colframe=memframe,
    colbacktitle=memtitlebg,
    coltitle=black,
    fonttitle=\bfseries\small,
    boxrule=0.6pt,
    arc=1.2mm,
    left=1.2mm,
    right=1.2mm,
    top=0.8mm,
    bottom=0.8mm,
    boxsep=0.8mm,
    before skip=4pt,
    after skip=4pt,
    attach boxed title to top left={xshift=2mm,yshift*=-2mm},
    boxed title style={boxrule=0pt,arc=1.2mm},
  }
}

\newcommand{\jsonkey}[1]{\texttt{#1}}
\newcommand{\jsonstr}[1]{\texttt{#1}}

\definecolor{gainred}{HTML}{C00000}
\newcommand{\up}[1]{\textcolor{gainred}{\ensuremath{\uparrow}\textbf{#1}}}

\newtcolorbox{promptbox}[1][]{
    skin=enhanced,
    colback=gray!2,        
    colframe=black!3,       
    arc=2pt,              
    title filled=true,
    coltitle=white,       
    colbacktitle=black!50,   
    fonttitle=\bfseries\Large,
    left=6pt,right=6pt,top=4pt,bottom=6pt,
    #1 
}

\def\BibTeX{{\rm B\kern-.05em{\sc i\kern-.025em b}\kern-.08em
    T\kern-.1667em\lower.7ex\hbox{E}\kern-.125emX}}
\begin{document}

\title{AdaRepair-Mem: Adaptive Experience Orchestration for Repository-Level Program Repair 
\author{
\large\bfseries
Zichen Luo\textsuperscript{1,\textdagger},
Jiachen Guo\textsuperscript{1,2,\textdagger},
Wenjun He\textsuperscript{1,2},
Siyu Wang\textsuperscript{1},
Jiongchi Yu\textsuperscript{1},
Fangming Zhao\textsuperscript{1},
\\
Yuan Chen\textsuperscript{2},
Te Cao\textsuperscript{2},
Liqun Liu\textsuperscript{2},
Ning Zheng\textsuperscript{2},
Wei Xu\textsuperscript{2},
Jie Jiang\textsuperscript{2},
Ziming Zhao \textsuperscript{1,*}
\\[8pt]
\normalfont\normalsize
\textsuperscript{1} Zhejiang University\qquad\textsuperscript{2} Tencent\\
}
\thanks{\textsuperscript{\textdagger}These authors contributed equally.}%
\thanks{\textsuperscript{*}Corresponding author.Email: zhaoziming@zju.edu.cn}%
}


\maketitle

\begin{abstract}
Recent memory-augmented repository-level program repair methods reuse historical repair experiences to improve LLM-based issue resolution. However, our analysis reveals three limitations in existing repository-level memory retrieval. First, episodic memory is highly imbalanced across repositories, leaving low-resource repositories with little effective support. Second, more memory does not monotonically lead to higher repair success, suggesting that relevance, quality, and redundancy matter more than raw memory volume. Third, memory accumulation is phase-misaligned: repositories may contain many reproduction experiences but few patch or refinement experiences. To address these problems, we propose an adaptive experience retrieval framework for repository-level program repair. Our framework introduces coverage-aware retrieval, which falls back to cross-repository or repair-type-based memories when same-repository memory is insufficient; quality-aware selection, which ranks memories by relevance, historical utility, specificity, and redundancy; and stage-aware routing, which separates and retrieves memories for reproduction, localization, patch generation, patch refinement, and validation. Evaluated on SWE-Bench-Lite and SWE-Bench-Verified, the proposed framework improves repair performance on under-covered repositories, reduces noisy memory retrieval, and better supports failed-to-fixed patch refinement. Our results show that the key to memory-augmented repair is not simply accumulating more experiences, but retrieving the right experiences for the right repair context.
\end{abstract}


\section{Introduction}
\label{sec:intro}
Repository-level automated program repair (APR) treats issue resolution as an end-to-end software engineering task: given a natural-language issue report and a code repository, the goal is to localize faulty code, generate a patch, and validate the fix against existing tests. The emergence of large language models (LLMs)~\cite{fan2023large,zheng2023survey,zan2023large,chen2021evaluating,li2022competition,austin2021program} and benchmarks such as SWE-Bench~\cite{jimenez2024swe,martinez2025dissecting} has driven rapid progress on this task, with recent agent-based and pipeline-based systems achieving increasingly strong results~\cite{xia2024agentless,wang2025openhands,ruan2025specrover,li2025patchpilot,aggarwal2025dars}.

Despite these advances, most non-memory systems solve each issue independently, without retaining knowledge acquired from previous repair attempts. Human developers, however, routinely draw on prior debugging and repair experience when addressing new issues. Motivated by this observation, recent work has introduced memory mechanisms into LLM agents~\cite{zhao2024expel,packer2023memgpt,zhong2024memorybank,chhikara2025mem0,anokhin2024arigraph,park2023generative,wang2023voyager,ferraz2026retrieval,du2026memory}, and ExpeRepair~\cite{mu2026experepair} has extended this idea to repository-level APR by storing repair demonstrations and distilled insights from prior trajectories. Its results show that historical repair experience can provide valuable guidance for issue reproduction and patch generation, establishing memory-augmented APR as a promising direction.
\begin{figure}[htbp]
    \centerline{\includegraphics[width=\linewidth]{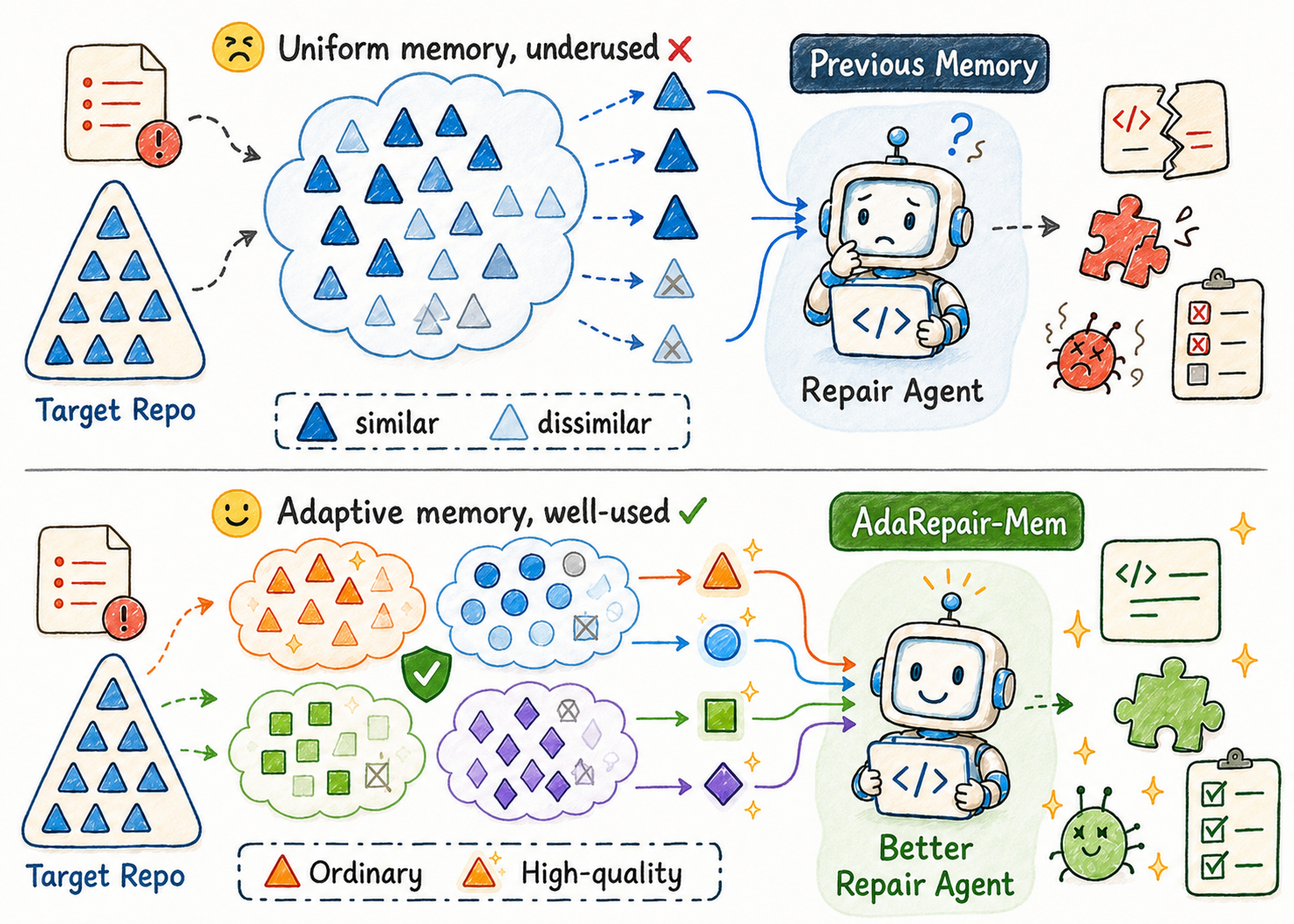}}
    \caption{Illustration comparing the traditional unified memory mechanism and our AdaRepair-Mem adaptive experience retrieval pipeline, which enables full utilization of high-value stage-matched repair memories.}
    \label{different-models}
\end{figure}

\begin{figure*}[htbp]
\centerline{\includegraphics[width=\linewidth]{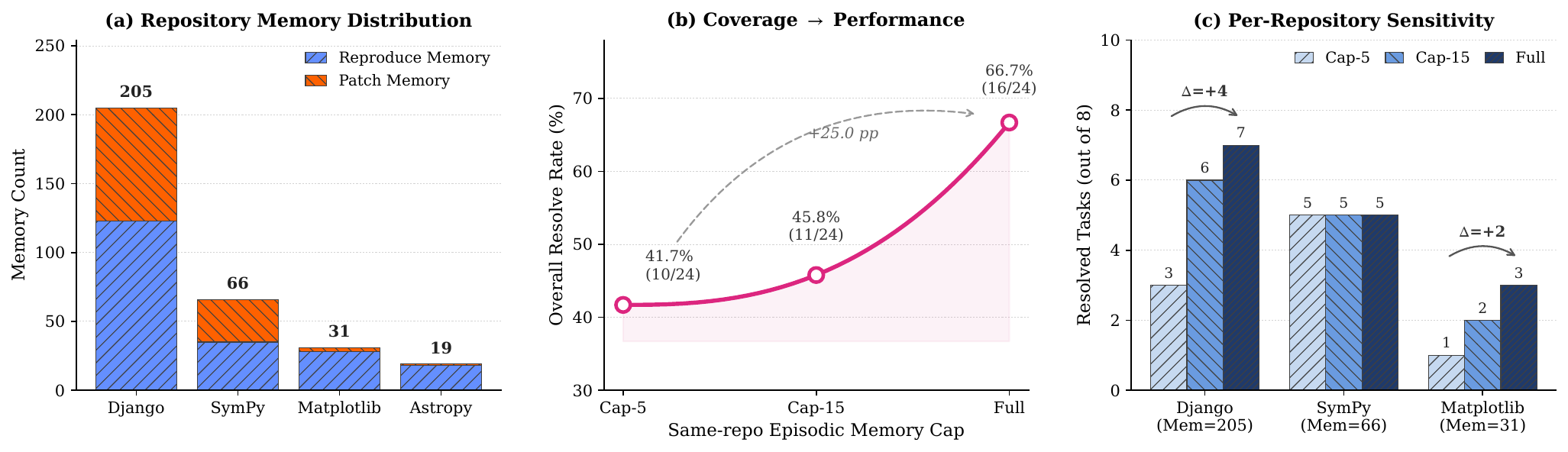}}
\caption{Motivational study of repository-level memory retrieval. (a) Imbalanced memory stock across different projects; (b) Repair performance grows with the upper bound of available same-repository memory; (c) Low-memory repositories are more sensitive to memory capacity restrictions, motivating our coverage-aware retrieval module.}
\label{moti}
\end{figure*}

While prior work has established the value of repair memories, how such memories should be organized and retrieved remains poorly understood~\cite{jiang2023active}. Existing approaches typically maintain memories within repository boundaries, retrieve them based primarily on lexical or semantic similarity~\cite{robertson2009probabilistic,liu2022makes,rubin2022learning}, and use a shared memory pool throughout the repair process. These design choices are largely heuristic, and their implications have not been systematically studied. Our analysis of repair memories in existing memory-augmented systems reveals several recurring limitations.
Repair memories are unevenly distributed across repositories. Some repositories accumulate a large number of repair experiences, while others contain only a few. Because retrieval is typically restricted to the current repository, repositories with limited historical data receive little benefit from memory augmentation. The effectiveness of memory-assisted repair therefore depends heavily on repository-specific memory availability~\cite{hassan2008road,chakraborty2025blaze,li2019deepfl,gao2023retrieval}.

Beyond coverage, a larger memory pool does not necessarily lead to better repair performance~\cite{liu2024lost,gao2023retrieval,jiang2023active,ferraz2026retrieval}. Our analysis shows that retrieval quality depends not only on similarity, but also on factors such as utility, specificity, and redundancy. Highly similar memories can still be unhelpful or misleading, while redundant memories consume context without providing additional guidance. Selecting useful memories is therefore at least as important as accumulating more of them.

Repair memories are also unevenly distributed across repair stages. In many repositories, reproduction-related experiences are abundant, whereas experiences for patch generation or patch refinement are scarce. Moreover, different stages require different forms of knowledge. For example, issue reproduction often relies on environment configuration and testing procedures, whereas patch refinement benefits from examples that capture how incorrect patches were revised into successful fixes. Using a single memory pool throughout the repair process therefore fails to account for the differing requirements of individual repair stages~\cite{li2025patchpilot,chen2025swe,shen2026structurally,pabba2025refine,zhang2026sgagent}.

These observations suggest that repository identity alone is insufficient for effective memory retrieval. Successful memory-augmented repair requires selecting experiences that are appropriate for the repository, the repair task, and the current stage of the repair process~\cite{gao2023retrieval,jiang2023active,ferraz2026retrieval,yao2022react,shen2026structurally}. Based on this insight, we propose AdaRepair-Mem, an adaptive experience retrieval framework for repository-level program repair.

AdaRepair-Mem addresses the limitations identified above through three complementary mechanisms. When repository-specific memories are insufficient, coverage-aware retrieval expands the search space to include related experiences from other repositories~\cite{hassan2008road,chakraborty2025blaze}. Quality-aware selection ranks candidate memories using multiple signals beyond similarity, including historical utility, specificity, and redundancy~\cite{gao2023retrieval,liu2024lost}. Stage-aware routing organizes memories according to different repair stages and retrieves stage-relevant experiences for reproduction, localization, patch generation, patch refinement, and validation~\cite{li2025patchpilot,shen2026structurally}. The framework is modular and can be integrated into existing memory-augmented repair systems without requiring changes to their underlying repair pipelines.
We evaluate AdaRepair-Mem on SWE-Bench-Lite and SWE-Bench-Verified~\cite{jimenez2024swe,martinez2025dissecting,badertdinov2026swe}. The results show that AdaRepair-Mem consistently improves repair performance on repositories with limited memory coverage, reduces ineffective memory retrieval, and achieves better token efficiency than the base system. 
Overall, we makes the following contributions:

\begin{figure*}[htbp]
\centerline{\includegraphics[width=\linewidth]{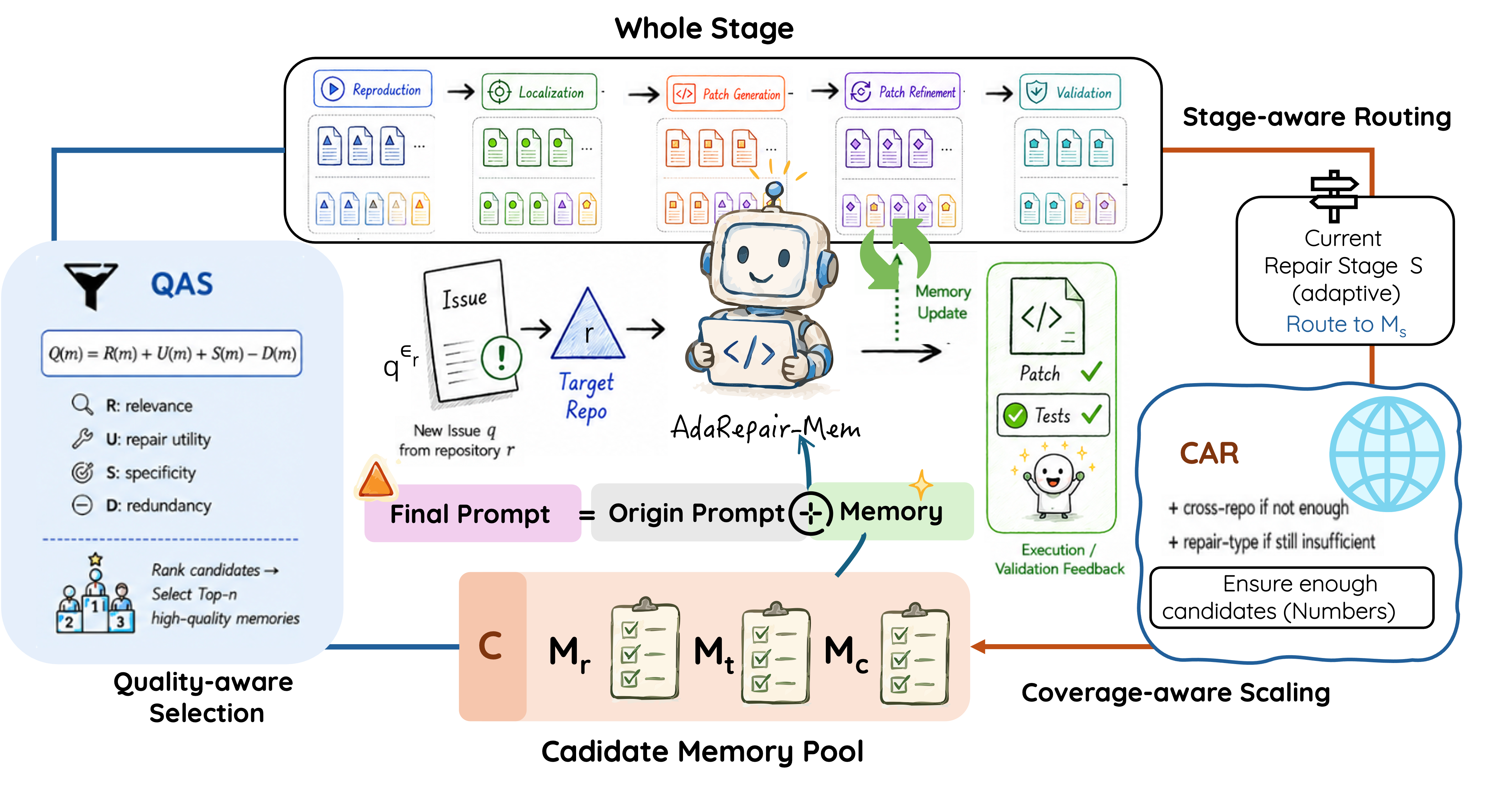}}
\caption{System overview of the proposed AdaRepair-Mem framework. Given a new bug issue from the target repository, the coverage-aware retrieval (CAR) module dynamically expands the candidate memory pool with cross-repository and same-type repair experiences when local memories are scarce. The quality-aware selection (QAS) module scores and filters high-value memories by jointly measuring relevance, repair utility, specificity and redundancy. Finally, the stage-aware routing (SAR) module distributes screened memories to the corresponding stage partition matching the current repair phase, and the selected stage-specific memories are injected into the original prompt to assist the repair agent. After patch execution and validation, new repair trajectories are updated back to the memory pool for future retrieval.}
\label{illustration}
\end{figure*}

\begin{itemize}
\item We analyze repair memories in repository-level APR and identify three limitations of existing memory retrieval strategies.

\item We propose AdaRepair-Mem, a novel adaptive experience retrieval framework for memory-augmented program repair.
Different from existing works that focus on expanding memory sources, our work innovates from the perspective of precise memory utilization: it unifies coverage-aware retrieval, quality-aware selection and stage-aware routing into one coherent pipeline, delivering high-value, stage-matched repair experiences.

\item We evaluate AdaRepair-Mem on SWE-Bench-Lite and SWE-Bench-Verified and show that it improves repair effectiveness, particularly for repositories with limited historical experiences, while achieving better token efficiency than the base system.

\end{itemize}

\section{Background and Motivation}
\label{sec:background_motivation}

\subsection{Repository-Level Program Repair}

Repository-level automated program repair (APR) resolves natural-language issues in the context of an entire software repository, requiring file localization, project-specific reasoning, patch generation, and validation~\cite{jimenez2024swe,xia2024agentless}. Recent LLM-based systems address this task through agentic interaction or staged repair pipelines~\cite{yang2024swe,zhang2024autocoderover,li2025patchpilot}, but most solve each issue independently and discard prior repair experience. Since human developers routinely reuse debugging knowledge from similar issues, memory-augmented repair stores historical trajectories and retrieves them as contextual guidance for future repairs~\cite{mu2026experepair,chen2025swe}.

\subsection{Repair Memories}

In this work, a repair memory denotes a structured record extracted from a previous repair attempt~\cite{mu2026experepair,chen2025swe}. A memory may include the issue description, retrieved code context, generated reproduction scripts, candidate patches, validation feedback, and the final successful repair artifact. These records are useful because they capture both project-specific knowledge and task-level repair patterns. For example, a reproduction memory may describe how to construct a minimal failing test under a specific project configuration, while a refinement memory may record how an initially incorrect patch was revised after receiving test feedback.
Existing memory-augmented repair systems typically organize these records by repository and retrieve memories using lexical or semantic similarity between the current issue and past repair cases~\cite{mu2026experepair,chen2025swe,robertson2009probabilistic}. This design is intuitive because same-repository memories often encode local APIs, naming conventions, and architectural assumptions. However, repository identity and surface-level similarity alone are insufficient to determine whether a memory is useful for the current repair context. A retrieved memory can be similar but unsuccessful, correct but too generic, or relevant to a different repair stage. As a result, memory retrieval should be viewed not merely as nearest-neighbor search, but as a policy for selecting which prior experiences should be exposed to the repair agent, from where they should be retrieved, and at which stage they should be used. Such a policy can prioritize memories that are not only similar to the present issue, but also demonstrably effective and actionable under comparable debugging conditions.

\subsection{Motivating Observations}

Our motivational study reveals three limitations in existing repository-level memory retrieval strategies.

\textbf{Observation 1: Repository-level memory coverage is imbalanced.} As shown in Fig.~\ref{moti}, repair memories are unevenly distributed across repositories. Some projects accumulate abundant historical repair experiences, whereas others contain only a small number of usable memories. A retrieval strategy restricted to the current repository therefore gives strong memory support to high-resource repositories but provides limited assistance to low-resource ones. This imbalance makes the effectiveness of memory-augmented repair depend heavily on repository-specific memory availability rather than on the intrinsic usefulness of historical repair knowledge.

\textbf{Observation 2: More memories do not necessarily imply better repair.} Increasing the memory pool can improve coverage, but it can also introduce noisy or redundant context. Similar memories may correspond to failed repair attempts, and multiple retrieved records may describe nearly identical repair steps. Since LLM repair agents operate under a limited context budget, injecting low-utility memories can distract the agent, waste tokens, and reduce the chance of generating a correct patch~\cite{liu2024lost,gao2023retrieval}. This suggests that retrieval quality is at least as important as memory quantity. A useful memory selection policy should consider not only relevance, but also historical repair utility, specificity of repair artifacts, and redundancy among retrieved records.

\textbf{Observation 3: Repair memories are stage-dependent.} Repository-level repair is a multi-stage process~\cite{li2025patchpilot,shen2026structurally}. Issue reproduction, fault localization, patch generation, patch refinement, and validation require different forms of historical knowledge. Reproduction often benefits from prior test construction and environment setup, localization benefits from examples of where similar symptoms were previously mapped in the codebase, and refinement benefits from failed-to-fixed patch trajectories. A unified memory pool ignores these stage-specific requirements and may deliver memories to stages where they provide limited actionable guidance. Therefore, effective memory retrieval should be aligned with the current repair stage.

\section{Method}

Fig.~\ref{illustration} illustrates the overall framework of AdaRepair-Mem, a novel adaptive memory construction and retrieval method for repository-level automatic program repair. Instead of adopting a simple unified memory pool and single similarity-based retrieval strategy, AdaRepair-Mem systematically redesigns the full pipeline of historical repair experience management. This design is motivated by three key observations from Section~\ref{sec:intro}: uneven distribution of repair memories across repositories, significant variance in memory quality, and stage-specific demands for repair experience.
Repository-level repair proceeds through five stages: issue reproduction, fault localization, patch generation, patch refinement, and validation. Before prompt construction for the repair agent at each stage, AdaRepair-Mem executes an adaptive retrieval module in three consecutive steps. The three modules jointly ensure \textbf{coverage adequacy of the candidate pool}, \textbf{memory quality of selected entries}, and \textbf{stage appropriateness of the final delivered context}.
The selected memories are injected into the current stage prompt without altering the core repair workflow. As a general retrieval enhancement framework, AdaRepair-Mem can be integrated into existing memory-augmented repair systems with minimal engineering overhead. We elaborate on the three components in the following subsections.

\subsection{Coverage-Aware Retrieval (CAR)}

Existing memory-augmented repair systems retrieve repair experiences only
from the current repository. Such a design implicitly assumes that every
repository has accumulated sufficient historical repair experiences.
However, as shown in Section~\ref{sec:intro}, repair memories are
highly imbalanced across repositories. While some repositories contain
abundant repair histories, others contain only a few usable memories.
Restricting retrieval to the current repository therefore limits the
available repair context for repositories with sparse historical
experience.

Coverage-aware Retrieval (CAR) addresses this problem by adaptively
expanding the retrieval scope according to memory availability. The
design follows a simple principle: repository-specific repair
experiences should always be preferred because they capture
project-specific implementation patterns and development conventions.
Broader repair knowledge is introduced only when repository-specific
memories are insufficient to provide adequate retrieval coverage.

Given an issue $q$ from repository $r$, let $M_r$ denote the repair
memories belonging to repository $r$, $M_c$ denote repair memories
retrieved from other repositories, and $M_t$ denote repair-type-based
memories that share similar repair patterns. In Algorithm~\ref{alg:car},
we additionally use $s$ to denote the current repair stage, $t$ to denote
the repair type of $q$, $\mathcal{M}$ to denote the constructed memory
bank, $B$ to denote the maximum number of cross-repository candidates,
and $A_s$ to denote the stage-specific admission rule. CAR progressively
constructs a candidate memory pool by expanding the retrieval scope until
a sufficient number of candidate memories is obtained. Unlike subsequent
retrieval stages, CAR does not determine which memories will ultimately
be used. Its objective is to construct a candidate pool with adequate
coverage for later quality-aware selection.
Formally, let $k$ denote the minimum number of candidate memories
required for subsequent retrieval. The candidate pool
$\mathcal{C}$ is constructed as

\begin{equation}
\mathcal{C}=
\begin{cases}
M_r,
&
|M_r|\ge k,
\\
M_r\cup M_c,
&
|M_r|<k
\land
|M_r\cup M_c|\ge k,
\\
M_r\cup M_c\cup M_t,
&
\text{otherwise}.
\end{cases}
\label{eq:car}
\end{equation}

This progressive expansion strategy preserves repository-specific repair knowledge whenever possible while avoiding the coverage limitations caused by repositories with sparse repair histories. The resulting candidate pool is subsequently passed to the Quality-aware Selection module for ranking and filtering.

Algorithm~\ref{alg:car} summarizes the overall retrieval procedure.
Repository-specific memories are first collected. If the candidate pool does not satisfy the retrieval threshold, CAR expands the search to cross-repository memories and then to repair-type-based memories. The resulting candidate pool is forwarded to the next retrieval stage.

\begin{algorithm}[t]
\caption{Coverage-Aware Retrieval (CAR)}
\label{alg:car}
\begin{algorithmic}[1]
\Require issue $q$, repository $r$, repair stage $s$, repair type $t$
\Require memory bank $\mathcal{M}$, coverage threshold $k$, global cap $B$, admission rule $A_s$
\Ensure candidate memory pool $\mathcal{C}$

\State $M_r \gets \textsc{LoadSameRepoMemory}(\mathcal{M}, r, s)$
\State $M_r \gets \textsc{FilterAdmitted}(M_r, A_s)$
\State $\mathcal{C} \gets M_r$

\If{$|\mathcal{C}| < k$}
    \State $M_c \gets \textsc{LoadCrossRepoMemory}(\mathcal{M}, r, s, B)$
    \State $M_c \gets \textsc{FilterAdmitted}(M_c, A_s)$
    \State $\mathcal{C} \gets \mathcal{C} \cup M_c$
\EndIf

\If{$|\mathcal{C}| < k$}
    \State $M_t \gets \textsc{LoadRepairTypeMemory}(\mathcal{M}, t, s, B)$
    \State $M_t \gets \textsc{FilterAdmitted}(M_t, A_s)$
    \State $\mathcal{C} \gets \mathcal{C} \cup M_t$
\EndIf

\State $\mathcal{C} \gets \textsc{RemoveDuplicateMemory}(\mathcal{C})$
\State \Return $\mathcal{C}$
\end{algorithmic}
\end{algorithm}

\subsection{Quality-Aware Selection (QAS)}

Coverage-aware Retrieval enlarges the candidate memory pool to improve
memory coverage. However, as shown in our empirical study, increasing the
number of retrieved memories alone does not necessarily improve repair
performance. Candidate memories often differ substantially in their
usefulness. Highly relevant memories may correspond to unsuccessful
repair attempts, while multiple retrieved memories may contain nearly
identical information or provide only generic repair guidance. Selecting
memories solely according to lexical or semantic similarity therefore
introduces noisy retrieval contexts and limits the effectiveness of
memory-augmented repair.

To address this problem, we propose \textbf{Quality-aware Selection
(QAS)}, which systematically evaluates the quality of candidate repair
memories using multiple complementary signals rather than similarity
alone. Given the candidate memory pool constructed by CAR, QAS estimates
the usefulness of each candidate memory from four perspectives:
\emph{relevance}, \emph{repair utility}, \emph{specificity}, and
\emph{redundancy}. These signals are lightweight to compute and are
directly derived from information already maintained during the repair
process.

\textbf{Relevance} measures how well a repair memory matches the current
repair context. Since different repair stages require different
information, QAS adopts stage-specific retrieval queries following the
original retrieval strategy of ExpeRepair. For example, patch generation
emphasizes the similarity between issue descriptions and historical
patches, whereas issue reproduction places greater emphasis on issue
descriptions and historical test cases. Relevance is computed using BM25~\cite{robertson2009probabilistic}
over the corresponding retrieval fields for each repair stage.

\textbf{Repair utility} reflects whether a repair memory corresponds to a
successful repair outcome. Rather than treating all memories equally,
QAS gives higher priority to memories that successfully reproduce issues,
generate valid patches, or pass subsequent verification. These signals
are directly obtained from the repair outcomes stored in each memory
entry.

\textbf{Specificity} estimates the amount of concrete repair information
contained in a memory. Memories with detailed issue descriptions, test
cases, or code patches generally provide richer repair context than short
or generic records. QAS therefore favors memories containing more
informative repair artifacts.

\textbf{Redundancy} measures the overlap between candidate memories.
QAS first removes exact duplicate memories and further filters highly
similar candidates using token-level Jaccard similarity, encouraging the
retrieved memories to provide complementary rather than repetitive
repair information.

The overall quality score of a candidate memory $m$ is computed as

\begin{equation}
Q(m)=R(m)+U(m)+S(m)-D(m),
\label{eq:qas}
\end{equation}

where $R(m)$, $U(m)$, $S(m)$, and $D(m)$ denote the relevance, repair
utility, specificity, and redundancy scores, respectively. Candidate
memories are ranked according to their quality scores, and the
highest-ranked memories are retained for subsequent stage-aware routing.

\subsection{Stage-Aware Routing (SAR)}

Coverage-aware Retrieval determines where candidate memories are
retrieved from, while Quality-aware Selection determines which memories
should be retained. The remaining question is when these memories should
be used during repository-level program repair. Existing memory-augmented
repair systems retrieve all memories from a unified memory store,
regardless of the repair stage in which they are consumed. However, our
empirical study shows that repair memories are unevenly distributed
across different repair stages, and memories that are useful for one
stage are often less informative for another.

To address this problem, we propose \textbf{Stage-aware Routing (SAR)},
which organizes repair memories according to the five repair stages of
ExpeRepair and dynamically routes stage-specific memories to the
corresponding repair stage. Rather than treating repair memory as a
single shared resource, SAR maintains separate memory partitions for
issue reproduction, fault localization, patch generation, patch
refinement, and patch validation. As the repair process proceeds, each
stage retrieves memories only from its corresponding memory partition.

Formally, let
$\mathcal{M}=\{M_{\text{rep}},M_{\text{loc}},M_{\text{gen}},M_{\text{ref}},M_{\text{val}}\}$
denote the five stage-specific memory partitions. Given the current
repair stage $s$, SAR routes the quality-ranked memories to the
corresponding stage by

\begin{equation}
M^{*}=f(s,\mathcal{M})=M_s,
\label{eq:sar}
\end{equation}

where $M_s$ denotes the memory partition associated with stage $s$. The
selected memories are then incorporated into the prompt of the current
repair stage.
\begin{table*}[htbp]
\centering
\caption{Main comparison results on SWE-Bench Lite and SWE-Bench Verified benchmarks. 
Baseline results are taken from the corresponding papers and the official SWE-Bench leaderboard.}
\label{tab:swe_main_compare}
\setlength{\tabcolsep}{4pt}
\renewcommand{\arraystretch}{1.03}
\footnotesize

\setlength{\aboverulesep}{0.2pt}
\setlength{\belowrulesep}{0.2pt}

\begin{tabular}{l c | c c | c c}
\toprule
\multirow[c]{2}{*}{\textbf{Method}}
& \multicolumn{1}{c|}{\multirow[c]{2}{*}{\textbf{LLM}}}
& \multicolumn{2}{c|}{\textbf{SWE-Bench Lite}}
& \multicolumn{2}{c}{\textbf{SWE-Bench Verified}} \\
\cmidrule(lr){3-4} \cmidrule(lr){5-6}
& \multicolumn{1}{c|}{}
& \textbf{pass@1 (\%)} & \textbf{AVG Cost (\$)}
& \textbf{pass@1 (\%)} & \textbf{AVG Cost (\$)} \\
\midrule
SWE-agent     & Claude 3.5 Sonnet               & 23.0 & 1.62  & 33.6 & 1.59 \\
Aider         & GPT-4o + Claude 3 Opus          & 26.3 & -     & -    & -    \\
AutoCodeRover & GPT-4o                          & 30.7 & -     & -    & -    \\
SpecRover     & Claude 3.5 Sonnet + GPT-4o      & 31.0 & \textbf{0.65}  & 46.2 & -    \\
Agentless-1.5 & Claude 3.5 Sonnet               & 40.7 & 1.12  & 50.8 & 1.19 \\
OpenHands     & CodeAct v2.1                    & 41.7 & 1.33  & 53.0 & \textbf{0.78} \\
PatchPilot    & Claude 3.5 Sonnet               & 45.3 & 0.97  & 53.6 & 0.99 \\
DARS          & Claude 3.5 Sonnet + DeepSeek R1 & 47.0 & 12.24 & -    & -    \\
ExpeRepair    & Claude 3.5 Sonnet + o4-mini     & 48.3 & 1.89  & 57.2 & 1.74 \\
\midrule
\rowcolor{gray!20}
\textbf{AdaRepair-Mem}
& Claude 3.5 Sonnet + o4-mini
& \textbf{51.0} & 1.67
& \textbf{63.2} & 1.52 \\
\bottomrule
\end{tabular}
\end{table*}
This stage-specific organization enables different repair stages to
focus on the repair experiences that are most relevant to their own
objectives. For example, issue reproduction primarily benefits from
historical test construction and environment configuration, whereas patch
generation relies more on successful code modifications, and patch
refinement benefits from failed-to-fixed repair trajectories. By routing
different memories to different repair stages, SAR reduces interference
between heterogeneous repair experiences and provides stage-appropriate
repair context throughout the repair pipeline.

Together, CAR, QAS, and SAR form a unified adaptive experience retrieval
framework. CAR expands the retrieval scope when repository-specific
memories are insufficient, QAS selects high-quality memories from the
candidate pool, and SAR delivers the selected memories to the repair
stage where they are most useful.
The framework integrates with ExpeRepair's repair pipeline, which includes test execution~\cite{ni2023lever} and validation~\cite{dinella2022toga} components to verify generated patches.









\section{Experiments}


\subsection{Setup}
\subsubsection{Benchmarks}
We evaluate AdaRepair-Mem on two widely used repository-level program repair benchmarks, SWE-Bench-Lite and SWE-Bench-Verified~\cite{jimenez2024swe}.
SWE-Bench-Lite contains 300 real GitHub issues sampled from popular Python repositories and has been widely used for evaluating LLM-based repair systems.
SWE-Bench-Verified contains 500 human-validated issues from SWE-Bench.
Compared with SWE-Bench-Lite, SWE-Bench-Verified provides a more reliable evaluation setting because each issue has been manually checked to ensure that the problem description is clear and that the associated test patch is suitable for judging correctness.
Recent analysis~\cite{badertdinov2026swe} has shown that repository-level benchmarks like SWE-Bench present substantially greater challenges than earlier function-level repair benchmarks~\cite{lu2021codexglue}, requiring deeper code understanding and multi-file reasoning.

\subsubsection{Baselines}
We compare AdaRepair-Mem with representative open-source repository-level program repair methods on SWE-Bench-Lite and SWE-Bench-Verified\footnote{\url{https://www.swebench.com/}}. These baselines are grouped into three categories according to their repair paradigm.

\textbf{Agent-based methods.}
These methods rely on autonomous tool use and iterative interaction with the repository.
SWE-agent~\cite{yang2024swe} equips an LLM with a general agent-computer interface for repository inspection, file editing, and test execution.
Aider~\cite{aider2026} is an interactive coding assistant that builds repository-aware context and performs Git-integrated code edits, following the paradigm of AI-assisted programming tools.
AutoCodeRover~\cite{zhang2024autocoderover} combines LLM reasoning with code search to localize and repair bugs in large repositories.
SpecRover~\cite{ruan2025specrover} extends this line by extracting specification and intent information from the codebase to improve autonomous repair.
OpenHands~\cite{wang2025openhands} is a general software engineering agent framework with sandboxed execution and solution ranking.
DARS~\cite{aggarwal2025dars} improves agentic repair through inference-time search, branching, and long-horizon feedback.

\textbf{Pipeline-based methods.}
These methods decompose repository-level repair into staged procedures rather than a fully autonomous agent loop.
Agentless~\cite{xia2024agentless} separates localization, patch generation, and validation in a procedural repair pipeline.
PatchPilot~\cite{li2025patchpilot} follows a five-stage workflow that includes reproduction, localization, patch generation, validation, and refinement.
\begin{figure}[htbp]
\centerline{\includegraphics[width=\linewidth]{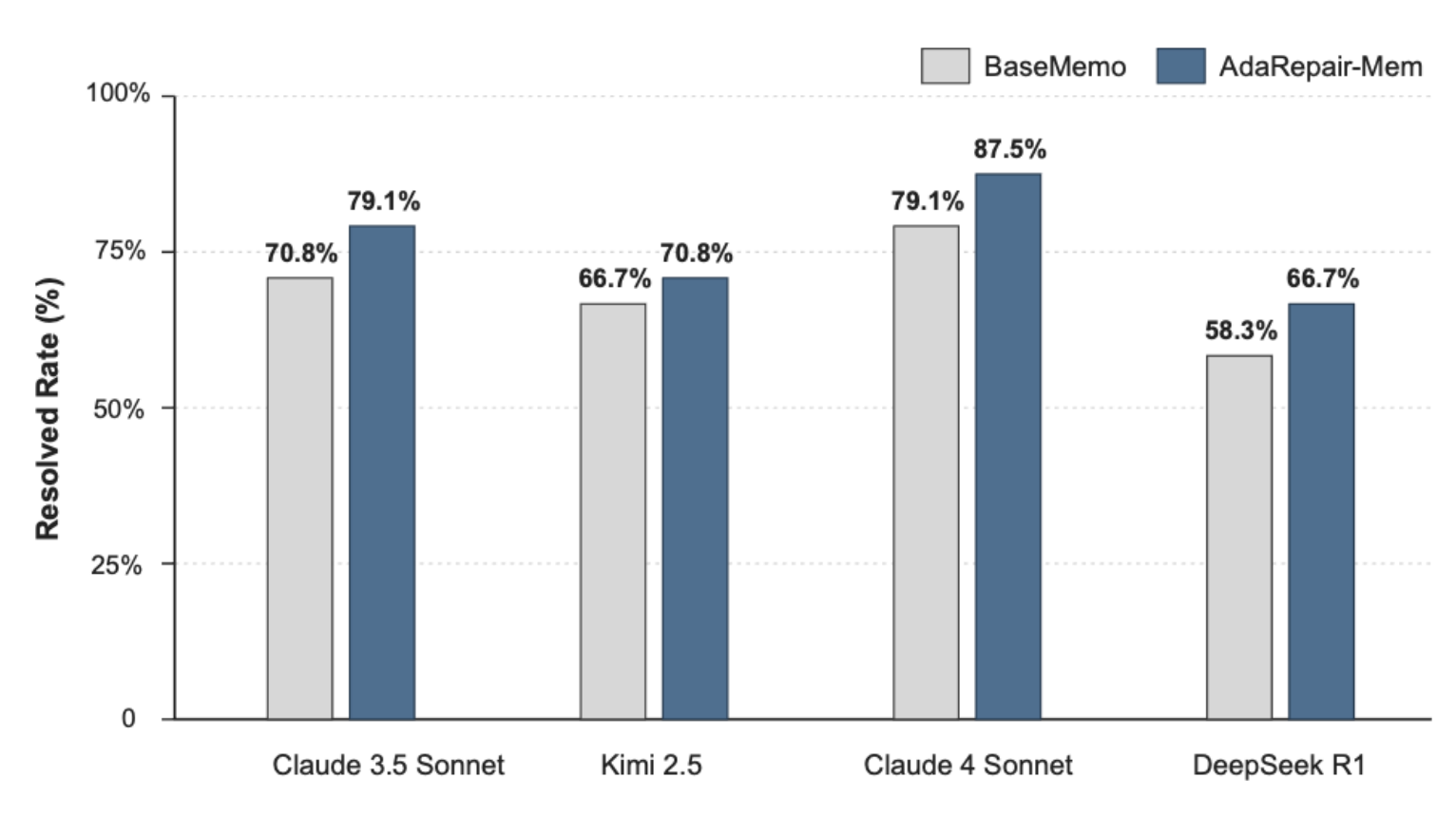}}
\caption{Comparison of previous memory-based repair and AdaRepair-Mem. AdaRepair-Mem adaptively retrieves memories by adaptive memory to provide more useful context for repair.}
\label{different-models}
\end{figure}
\begin{table*}[htbp]
\centering
\caption{File localization results comparison between RepoRepair, Agentless Lite, and AdaRepair-Mem on SWE-Bench Lite.\\
AL = Agentless Lite; RR = RepoRepair; Ada-Mem = AdaRepair-Mem.}
\label{tab:file_localization_results}
\setlength{\tabcolsep}{9pt}
\renewcommand{\arraystretch}{1.03}
\footnotesize

\setlength{\aboverulesep}{0.2pt}
\setlength{\belowrulesep}{0.2pt}

\begin{tabular}{c c|ccc|ccc}
\toprule
\multirow[c]{2}{*}{\bfseries Repository}
& \multirow[c]{2}{*}{\bfseries Instances}
& \multicolumn{3}{c|}{\bfseries File Retrieval (\%)}
& \multicolumn{3}{c}{\bfseries File Localization (\%)} \\
\cmidrule(lr){3-5} \cmidrule(lr){6-8}
& & \textbf{AL} & \textbf{RR} & \textbf{Ada-Mem}
& \textbf{AL} & \textbf{RR} & \textbf{Ada-Mem} \\
\midrule
astropy/astropy           & 6   & 100   & 100   & 100   & 83.33 & 100   & 100 \\
django/django             & 114 & 95.61 & 96.49 & 96.49 & 85.09 & 87.72 & 92.98 \\
matplotlib/matplotlib     & 23  & 91.30 & 95.65 & 95.65 & 43.48 & 69.57 & 95.65 \\
mwaskom/seaborn           & 4   & 100   & 100   & 100   & 50.00 & 75.00 & 100 \\
pallets/flask             & 3   & 100   & 100   & 100   & 100   & 100   & 100 \\
psf/requests              & 6   & 100   & 100   & 83.33 & 83.33 & 100   & 83.33 \\
pydata/xarray             & 5   & 100   & 100   & 100   & 80.00 & 100   & 100 \\
pylint-dev/pylint         & 6   & 100   & 66.67 & 83.33 & 100   & 50.00 & 83.33 \\
pytest-dev/pytest         & 17  & 100   & 88.24 & 94.12 & 94.12 & 5.88  & 100 \\
scikit-learn/scikit-learn & 23  & 100   & 95.65 & 95.65 & 65.22 & 73.91 & 95.65 \\
sphinx-doc/sphinx         & 16  & 100   & 100   & 100   & 75.00 & 75.00 & 100 \\
sympy/sympy               & 77  & 94.81 & 89.61 & 88.31 & 63.64 & 84.42 & 87.01 \\
\midrule
\bfseries Total
& \bfseries 300
& \bfseries 96.33
& 94.00
& 94.00
& 74.67
& 79.00
& \bfseries 92.67 \\
\bottomrule
\end{tabular}
\end{table*}
\textbf{Memory-augmented repair.}
This category reuses historical repair experiences to support new issue resolution.
ExpeRepair~\cite{mu2026experepair} is the closest baseline to our work because it augments repository-level repair with episodic and semantic memories.
Compared with ExpeRepair, AdaRepair-Mem keeps the same repair setting but replaces its retrieval strategy with coverage-aware retrieval, quality-aware selection, and stage-aware routing.

\subsubsection{Metrics}
Following ExpeRepair and recent repository-level repair systems~\cite{jimenez2024swe,xia2024agentless,mu2026experepair}, we report the following metrics. 

\begin{itemize}
    \item \textbf{Pass@1}: The percentage of issues resolved on the first attempt.
    \item \textbf{Average Cost}: The average inference cost per issue, measured in USD.
    \item \textbf{Token Efficiency}: The number of resolved tasks per 1,000 input tokens, where higher values indicate better use of the context budget.
    \item \textbf{Execution Success Rate (ESR)}: The percentage of generated reproduction scripts that execute successfully without errors, such as missing dependencies or configuration issues.
\end{itemize}

\subsubsection{Implementation Details}
We follow ExpeRepair's memory construction mechanism and repair pipeline, replacing only its retrieval strategy.
For coverage-aware retrieval (CAR), we set the initial same-repository retrieval target to $k_{\text{same}}=10$ and expand to cross-repository and repair-type pools when fewer than $k_{\text{same}}$ candidates are available, with cross-repository pool size capped at $k_{\text{cross}}=50$ to balance coverage and noise.
For quality-aware selection (QAS), we rank candidate memories by a weighted combination of relevance (BM25 score), repair utility (success indicator), specificity (trajectory length), and redundancy (pairwise similarity), and select the top-3 highest-ranked memories for each repair stage.
The quality weights prioritize relevance and utility (0.8 total) over specificity and redundancy (0.2 total), reflecting the observation that task alignment is more predictive than structural diversity.
For stage-aware routing (SAR), memories are partitioned into five stage-specific pools based on the stage labels recorded during ExpeRepair's memory construction.
These design choices prioritize retrieval precision over coverage and were validated on a small held-out set of 24 issues (6 from each coverage group) before being applied to the main evaluation.
\begin{table}[t]
\centering
\caption{Repair performance across repositories with different memory coverage levels. 
Cover: memory coverage level; Memo: the number of repair memories per repository.}
\label{tab:coverage_levels}
\setlength{\tabcolsep}{5.0pt}
\renewcommand{\arraystretch}{1.05}
\footnotesize

\setlength{\aboverulesep}{1.5pt}
\setlength{\belowrulesep}{1.5pt}

\begin{tabular}{c|cc|cc}
\toprule
\textbf{Cover} & \textbf{Repo} & \textbf{Memo} & \textbf{BaseMemo} & \textbf{AdaRepair-Mem} \\
\midrule
\multirow[c]{2}{*}{High}
& django & 82 & \multirow[c]{2}{*}{63.1\%} & \multirow[c]{2}{*}{\textbf{65.8\%}~\up{2.7}} \\
& sympy  & 31 & & \\
\midrule
\multirow[c]{2}{*}{Medium}
& flask  & 18 & \multirow[c]{2}{*}{41.2\%} & \multirow[c]{2}{*}{\textbf{52.3\%}~\up{11.1}} \\
& sphinx & 15 & & \\
\midrule
\multirow[c]{2}{*}{Low}
& astropy    & 3 & \multirow[c]{2}{*}{55.3\%} & \multirow[c]{2}{*}{\textbf{58.3\%}~\up{3.0}} \\
& matplotlib & 1 & & \\
\bottomrule
\end{tabular}
\end{table}

\subsection{Results}
\subsubsection{Overall Performance}
Table~\ref{tab:swe_main_compare} reports the main comparison on SWE-Bench-Lite and SWE-Bench-Verified.
Compared with ExpeRepair under the same Claude 3.5 Sonnet + o4-mini setting, AdaRepair-Mem improves pass@1 from 57.2\% to 63.2\%, a gain of 6.0 percentage points, while the average cost decreases from \$1.74 to \$1.52. The bottleneck of memory-augmented repair is not only whether historical experiences are available, but whether the system can select suitable experiences under repository coverage, memory quality, and repair-stage constraints. In other words, adaptive retrieval makes repair knowledge more effective by reducing irrelevant or poorly matched memories before they enter the repair context. If the gain came mainly from more exploration or larger prompts, we would expect higher inference cost. Instead, AdaRepair-Mem obtains higher pass@1 with lower average cost, indicating that the retrieved memories help the repair agent reach useful repair actions with less wasted context and fewer ineffective repair steps. Compared with agent-based and pipeline-based baselines, this shows that improving the memory policy can provide a complementary source of progress without changing the underlying repair workflow.

\subsubsection{Ablation Study}
For the ablation study, we use a fixed 24-instance diagnostic subset from SWE-Bench-Verified, covering BaseMemo-resolved, AdaRepair-Mem-resolved, BaseMemo-unresolved, and AdaRepair-Mem unresolved cases with six instances each. All variants are evaluated on the same subset, enabling a paired comparison of CAR, QAS, and SAR under both successful and failed repair trajectories.

Table~\ref{tab:ablation_memory} shows that all three modules are effective and complementary. On the 24-instance diagnostic subset, removing QAS causes the largest drop in pass@1, indicating that selecting high-quality memories is more important than retrieving more memories. Removing SAR reduces both pass@1 and ESR and increases the average cost, showing that stage-aware routing helps the system reuse memories at the right repair phase. Overall, the full AdaRepair-Mem achieves the best accuracy and the lowest cost, confirming that the gains come from the joint effect of CAR, QAS, and SAR rather than from any single component.

\subsubsection{Repository Coverage Analysis}
For this analysis, we use a fixed 150-instance stratified subset from SWE-Bench-Verified. The subset is constructed only for coverage-oriented analysis, while the main comparison is still conducted on the full benchmark.  We group repositories by the number of available repair memories before evaluation and select six repositories from three coverage levels: django and sympy for high coverage, flask and sphinx for medium coverage, and astropy and matplotlib for low coverage.  All compared methods are evaluated on the same sampled instances, and the grouping is based on memory availability rather than repair outcomes. This subset is used only for coverage-oriented and token-efficiency analyses. 
\begin{figure}[t]
\centerline{\includegraphics[width=\linewidth]{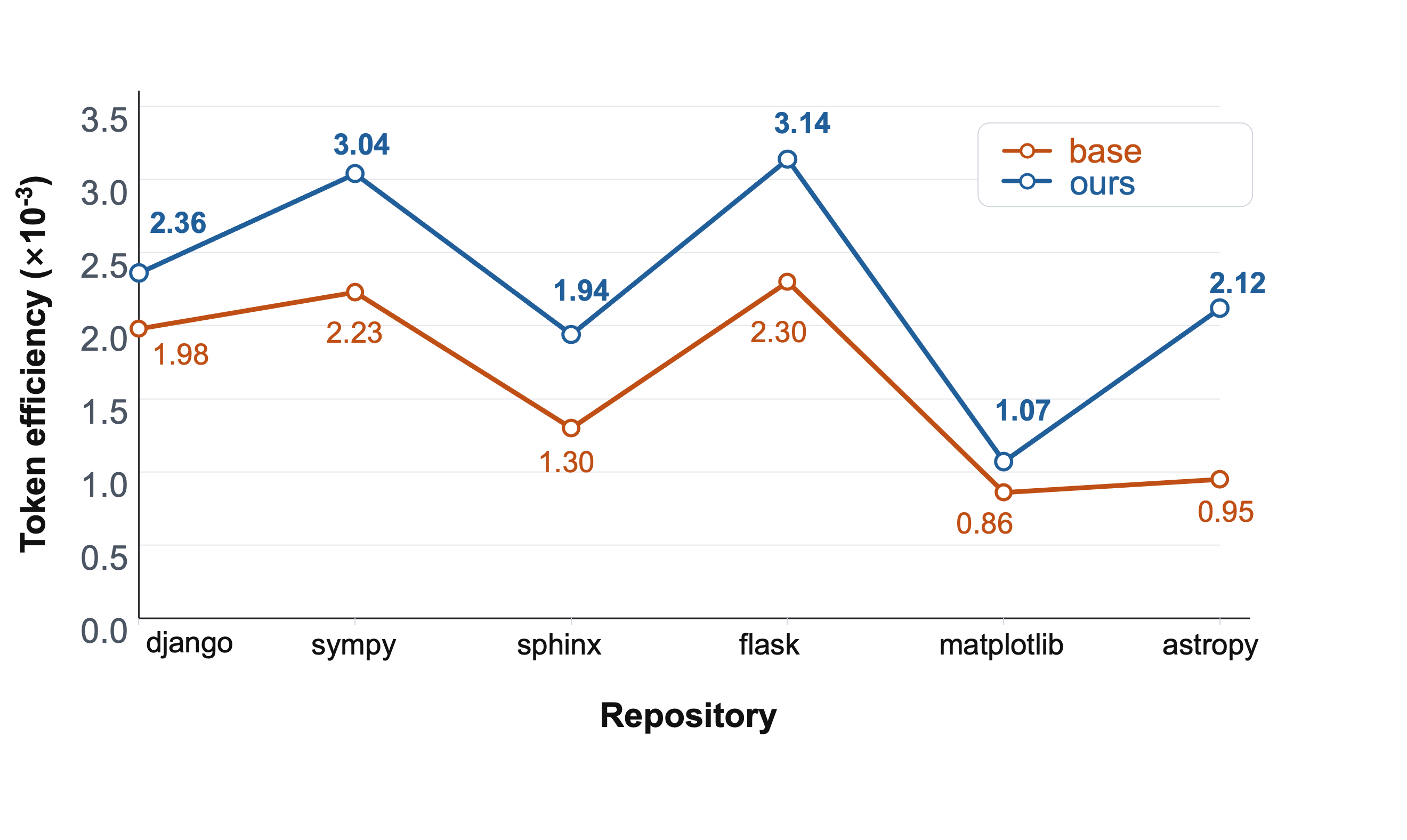}}
\caption{Token Efficiency Across Repositories: resolved tasks per 1,000 API input tokens; higher is better.
}
\label{token-effi}
\end{figure}
Table~\ref{tab:coverage_levels} shows that the benefit of AdaRepair-Mem is most visible when repository-local memory is incomplete. 
On the high-coverage repositories, AdaRepair-Mem improves pass@1 from 63.1\% to 65.8\%, suggesting that coverage-aware retrieval does not hurt repositories that already have sufficient local experience. 
The largest gain appears on the medium-coverage repositories, where pass@1 increases from 41.2\% to 52.3\%. This indicates that adaptive retrieval is especially useful when repository-local memories exist but do not fully cover the current repair pattern. 
Even on the low-coverage repositories, where the local memory pool is extremely small, AdaRepair-Mem still raises pass@1 from 55.3\% to 58.3\%. Taken together, these results show that repair performance is not determined by memory quantity alone, but by whether the retrieved experiences match the current repository and repair need.

\begin{table*}[htbp]
\centering
\setlength{\tabcolsep}{3.8pt} 
\caption{Ablation study on memory modules of AdaRepair-Mem. }
\footnotesize
\begin{tabular}{l c c c | c c c }
\toprule
\multirow{2}{*}{\bfseries Method} & \multicolumn{3}{c|}{\bfseries SWE-Bench Lite} & \multicolumn{3}{c}{\bfseries SWE-Bench Verified} \\
\cline{2-4} \cline{5-7}  
& \textbf{pass@1 (\%)} & \textbf{ESR (\%)} & \textbf{AVG Cost(\$)}  & \textbf{pass@1 (\%)} & \textbf{ESR (\%)} & \textbf{AVG Cost(\$)} \\
\midrule
\rowcolor{gray!20}\bfseries
\quad AdaRepair-Mem & 75.0 & 75.0& 2.55&70.8& 75.0&2.18\\
\quad w/o CAR & 75.0 &   62.5& 2.43 & 50.0 &  62.5& 2.13 \\
\quad w/o QAS & 58.3 &  50.0& 2.70& 45.8 &75.0& 2.54  \\
\quad w/o SAR & 66.7 & 45.8& 2.70& 62.5 & 66.7&  2.39 \\
\quad baseline &  45.8&  45.8& 2.91& 45.8& 50.0&2.78\\

\bottomrule
\end{tabular}
\label{tab:ablation_memory}
\end{table*}

\subsubsection{Token Efficiency}
Fig.~\ref{token-effi} further shows that AdaRepair-Mem uses the context budget more effectively across all six repositories. Compared with the base system, AdaRepair-Mem achieves higher token efficiency on every repository, with particularly clear gains on astropy, sympy, and flask. For example, token efficiency improves from 0.95 to 2.12 on astropy, from 2.23 to 3.04 on sympy, and from 2.30 to 3.14 on flask. These gains indicate that the improvement does not come from simply injecting more memory into the prompt. Instead, CAR and QAS help filter and expand memory more selectively, so that each 1,000 input tokens carries more repair-relevant information.
This is consistent with our claim that adaptive retrieval improves the utility of the context budget rather than increasing its size.

\subsubsection{Generalization across LLMs}
For the model generalization study, we use a fixed 24-instance subset from SWE-Bench-Verified, with eight cases sampled from each memory coverage level.  This balanced design avoids over-representing a single coverage group and allows us to compare different backbone models under the same set of issues. The results on different backbone models show that AdaRepair-Mem is not tied to a specific LLM.
Across Kimi 2.5~\cite{team2026kimi}, Claude 4 Sonnet~\cite{anthropic2025claude4}, Claude 3.5 Sonnet~\cite{anthropic2024claude3.5}, and DeepSeek R1~\cite{guo2025deepseek}, the bar chart consistently favors AdaRepair-Mem over the BaseMemo. This suggests that the proposed retrieval strategy is model-agnostic and can benefit both stronger and weaker reasoning models. The gain comes from better memory selection rather than from a particular prompt style or backbone capability. The generalization result supports our design goal of making adaptive experience retrieval a reusable plug-in for different repository-level repair systems.
\begin{figure}[t]
    \centering
    \includegraphics[width=\linewidth]{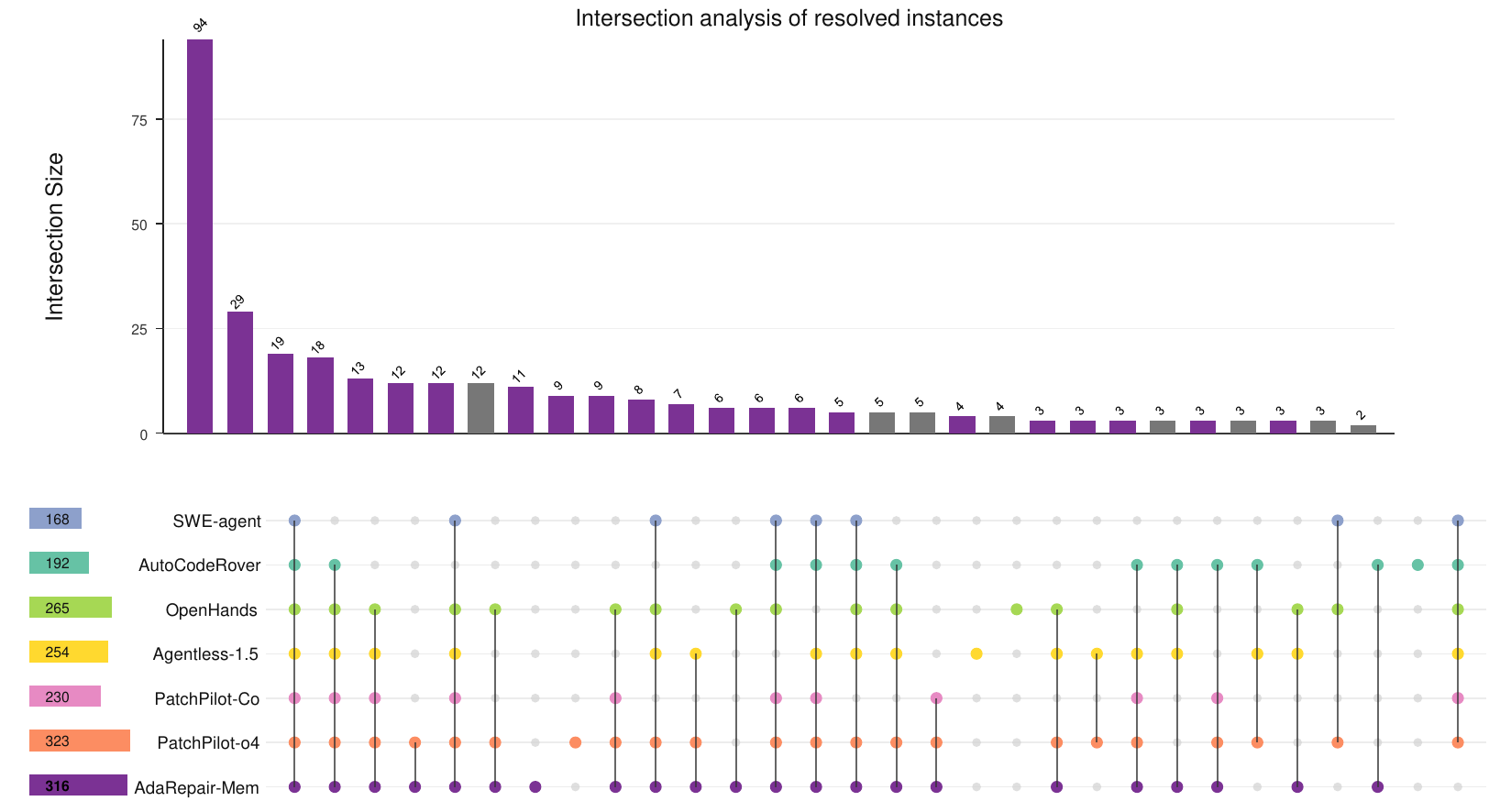}
\caption{Intersection analysis of resolved instances across seven repository-level repair systems. Purple bars indicate intersections that include AdaRepair-Mem.}
    \label{fig:intersection_analysis}
\end{figure}

\subsubsection{Cumulative resolution rates}
\label{fig:repo_cumulative_all}
Fig.~\ref{fig:repo_cumulative_all} reports the cumulative resolution rates of AdaRepair-Mem across 12 Python repositories in SWE-Bench-Verified. 
The repositories cover large, medium-size, and long-tail projects with diverse code structures and memory coverage levels. The curves show clear differences in repository-level repair difficulty, but AdaRepair-Mem maintains stable behavior across these settings. 
On the largest repository, django, the cumulative rate stabilizes after about 50 processed issues and stays within a narrow range without sustained degradation. SymPy maintains a resolution rate above 60\% for most of the process, and scikit-learn also remains at a relatively high level despite having fewer instances. For more challenging or lower-coverage repositories such as astropy and matplotlib, the curves fluctuate more because of smaller sample sizes, but they still converge to non-trivial repair rates rather than collapsing. 
Together with Table~\ref{tab:coverage_levels}, this result supports RQ1 by showing that AdaRepair-Mem can mitigate repository-level memory imbalance and provide robust repair behavior across repositories of different sizes and difficulties.

\begin{figure}[htbp]
\centerline{\includegraphics[width=\linewidth]{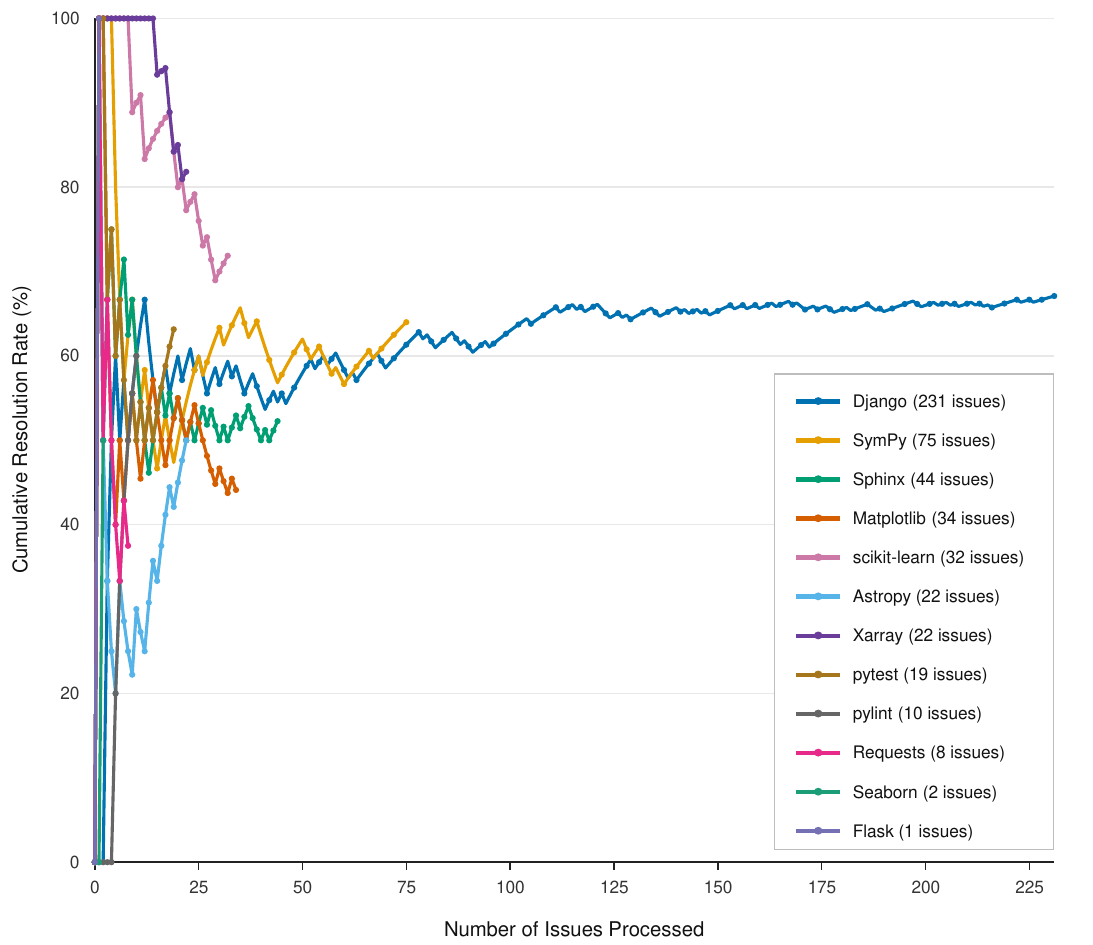}}
\caption{Cumulative resolution rates of AdaRepair-Mem across repositories on SWE-Bench-Verified.}
\label{fig:repo_cumulative_all}
\end{figure}

\begin{figure}[!t]
\centering

\begin{tcolorbox}[memorycard,title={Reproduce Memory}]
{\ttfamily\footnotesize
\{
\jsonkey{memory\_id}: \jsonstr{sphinx::M\_rep},

\jsonkey{repair\_stage}: \jsonstr{M\_rep},

\jsonkey{issue}: \jsonstr{Decorated \_\_init\_\_ missing in docs},

\textcolor{ctxgreen}{\jsonkey{retrieved\_context}: \jsonstr{Previous reproduction template}},

\jsonkey{generated\_test\_or\_patch}: \jsonstr{Minimal reproduction script},

\textcolor{feedred}{\jsonkey{failure\_feedback}: \jsonstr{Build path is incorrect}},

\jsonkey{successful\_output}: \jsonstr{Reusable reproduction harness}
\}
}
\end{tcolorbox}

\begin{tcolorbox}[memorycard,title={Localize Memory}]
{\ttfamily\footnotesize
\{
\jsonkey{memory\_id}: \jsonstr{sphinx::M\_loc},

\jsonkey{repair\_stage}: \jsonstr{M\_loc},

\jsonkey{issue}: \jsonstr{HTTP error masked by anchor check},

\textcolor{ctxgreen}{\jsonkey{retrieved\_context}: \jsonstr{Candidate source files}},

\jsonkey{generated\_test\_or\_patch}: \jsonstr{Bug localized to check\_thread()},

\textcolor{feedred}{\jsonkey{failure\_feedback}: \jsonstr{Initial search too broad}},

\jsonkey{successful\_output}: \jsonstr{Anchor-handling branch identified}
\}
}
\end{tcolorbox}

\begin{tcolorbox}[memorycard,title={Generate Memory}]
{\ttfamily\footnotesize
\{
\jsonkey{memory\_id}: \jsonstr{requests::M\_gen},

\jsonkey{repair\_stage}: \jsonstr{M\_gen},

\jsonkey{issue}: \jsonstr{GET sends Content-Length},

\textcolor{ctxgreen}{\jsonkey{retrieved\_context}: \jsonstr{Similar protocol header fix}},

\jsonkey{generated\_test\_or\_patch}: \jsonstr{Skip Content-Length for safe methods},

\textcolor{feedred}{\jsonkey{failure\_feedback}: \jsonstr{Header emitted unnecessarily}},

\jsonkey{successful\_output}: \jsonstr{Emit only when body exists}
\}
}
\end{tcolorbox}

\begin{tcolorbox}[memorycard,title={Patch-refinement Memory}]
{\ttfamily\footnotesize
\{
\jsonkey{memory\_id}: \jsonstr{pytest::M\_ref},

\jsonkey{repair\_stage}: \jsonstr{M\_ref},

\jsonkey{issue}: \jsonstr{caplog.clear breaks synchronization},

\textcolor{ctxgreen}{\jsonkey{retrieved\_context}: \jsonstr{Previous failed patch}},

\jsonkey{generated\_test\_or\_patch}: \jsonstr{Clear shared list in place},

\textcolor{feedred}{\jsonkey{failure\_feedback}: \jsonstr{Replacing list breaks alias}},

\jsonkey{successful\_output}: \jsonstr{Preserve shared record list}
\}
}
\end{tcolorbox}

\begin{tcolorbox}[memorycard,title={Validation Memory}]
{\ttfamily\footnotesize
\{
\jsonkey{memory\_id}: \jsonstr{sphinx::M\_val},

\jsonkey{repair\_stage}: \jsonstr{M\_val},

\jsonkey{issue}: \jsonstr{Whitespace in LaTeX inline code},

\textcolor{ctxgreen}{\jsonkey{retrieved\_context}: \jsonstr{Previous validation case}},

\jsonkey{generated\_test\_or\_patch}: \jsonstr{Validate PDF inline rendering},

\textcolor{feedred}{\jsonkey{failure\_feedback}: \jsonstr{Regression detected in PDF output}},

\jsonkey{successful\_output}: \jsonstr{No regression in inline code rendering}
\}
}
\end{tcolorbox}

\caption{Representative memory records from the stage-specific memory banks of AdaRepair-Mem. Retrieved context is highlighted in green and failure feedback in red.}
\label{fig:memory_cards}
\end{figure}

\begin{figure}[htbp]
\centering
\includegraphics[width=\linewidth]{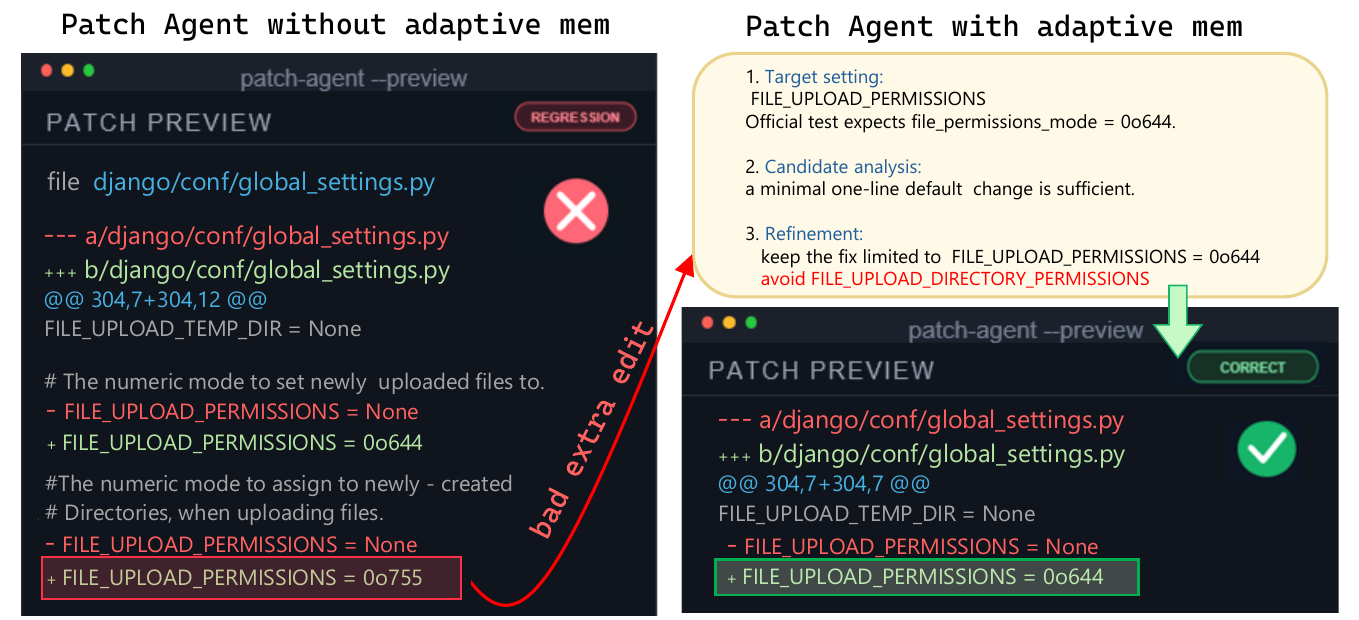}

\includegraphics[width=\linewidth]{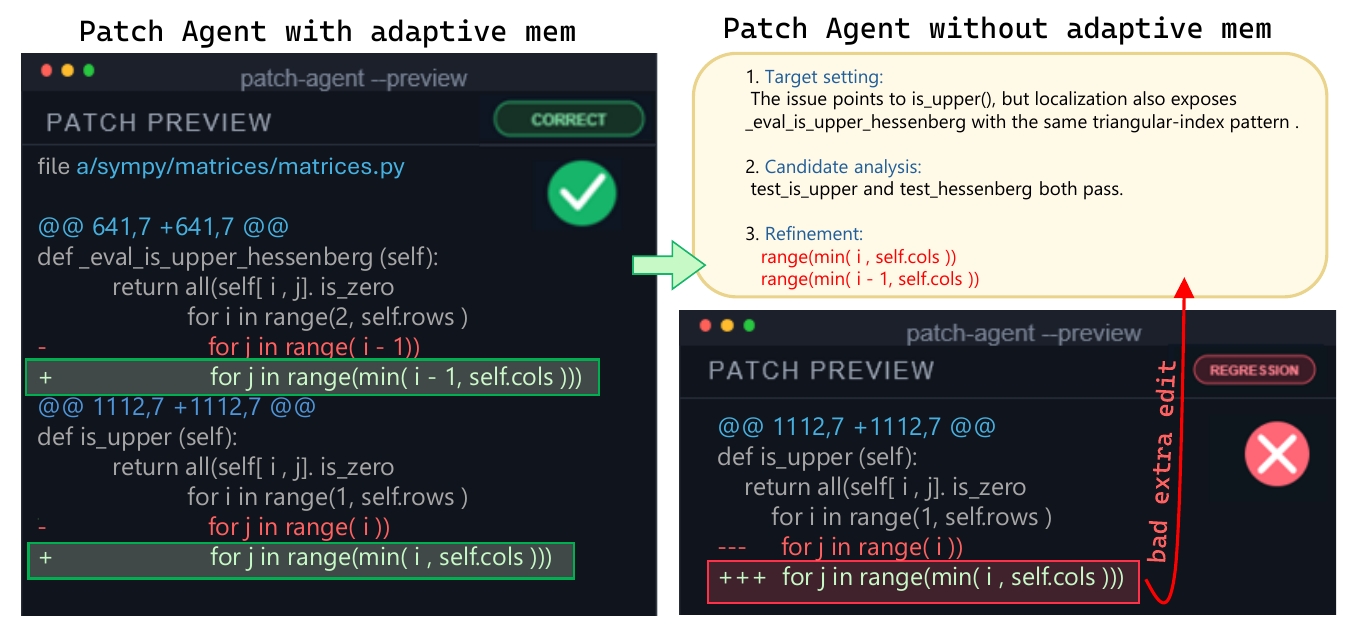}
\caption{Two real cases from the SymPy and Django repositories resolved by AdaRepair-Mem.}
\label{fig:case1}
\end{figure}

\subsubsection{File Localization Analysis}

To further understand where the improvement of AdaRepair-Mem originates, we isolate its file-level localization performance, a critical prerequisite for successful patch generation.
We compare AdaRepair-Mem with two systems that adopt different context strategies for this stage: Agentless Lite~\cite{xia2024agentless}, a lightweight variant of Agentless that relies on embedding-based retrieval and syntactic filtering; and RepoRepair~\cite{pan2026reporepair}, which leverages hierarchically generated code documentation to guide fault localization.

Table~\ref{tab:file_localization_results} reports per-repository results on SWE-Bench Lite.
For file retrieval, all three systems perform comparably (94\%--96\%), indicating that initial candidate identification does not meaningfully differentiate them.
For file localization, however, AdaRepair-Mem achieves 92.67\% accuracy, substantially outperforming RepoRepair (79.00\%) and Agentless Lite (74.67\%).
The improvement is most pronounced on repositories where historical repair experiences provide rich structural cues: AdaRepair-Mem correctly localizes all 17 pytest issues (compared to 1 for RepoRepair) and 22 of 23 matplotlib issues (compared to 10 for Agentless Lite).

These results suggest that adaptively retrieved repair memories deliver more actionable localization guidance than generic code documentation or embedding-based retrieval.
This finding aligns with the design of Stage-aware Routing (SAR): by routing stage-specific repair experiences to the localization stage, AdaRepair-Mem supplies the agent with historical examples showing where similar issues were previously localized, complementing lexical and documentation-based similarity signals.
\subsubsection{Intersection analysis.}
Fig.~\ref{fig:intersection_analysis} analyzes how the resolved instances of AdaRepair-Mem overlap with six representative baselines. The seven systems jointly resolve 393 unique instances, and 94 instances are solved by all methods. Beyond this common subset, many large intersections still include AdaRepair-Mem, indicating that AdaRepair-Mem remains competitive on broadly solvable issues while also contributing complementary repaired cases. The remaining 73 instances belong to smaller intersection groups that are omitted from the main bars for readability. This result suggests that adaptive memory retrieval improves repair coverage beyond simply reproducing the behavior of existing agentic repair systems.

\subsubsection{Stage-Specific Memory Records}
Figure~\ref{fig:memory_cards} shows representative memories from the stage-specific memory banks in AdaRepair-Mem. Each memory stores the issue, retrieved context, generated artifact, failure feedback, and successful output, preserving both useful experience and repair signals. Among these fields, retrieved context and failure feedback are the most informative for future retrieval, while the stage-wise organization provides the basis for SAR and helps CAR and QAS select more relevant memories.

\section{Related Work}

\subsection{Repository-Level Program Repair}
Large language models have advanced automated program repair (APR) from function-level patch generation to repository-level issue resolution~\cite{jimenez2024swe,fan2023large,zheng2023survey}. Existing methods mainly follow two paradigms: agent-based approaches that enable LLMs to interact with repositories via tool use and iterative reasoning~\cite{yang2024swe,zhang2024autocoderover,antoniades2025swesearch}, and pipeline-based approaches that decompose repair into staged workflows of reproduction, localization and validation~\cite{xia2024agentless,li2025patchpilot}. Despite differing execution paradigms, these methods generally treat each issue as an independent task and do not retain or adaptively reuse historical repair experience across tasks.

\subsection{Memory-Augmented Program Repair}
To enable cross-task knowledge reuse, memory mechanisms have been gradually introduced into APR. Early learning-based and template-based methods~\cite{jiang2021cure,liu2019tbar,bader2019getafixlearningfixbugs} first demonstrated that repair knowledge can be extracted from historical fixes and reusable patterns. Recent retrieval-augmented works~\cite{yang2025rag,mansur2024ragfix} further show that external code and patch context can improve LLM-based repair quality. Representative repository-level memory systems, including HAFix~\cite{shi2026hafix}, SWE-Exp~\cite{chen2025swe}, ExpeRepair~\cite{mu2026experepair} and ConRAD~\cite{li2026conrad}, have validated the value of historical repair experience. However, they all adopt component-specific retrieval designs, and no prior work formulates repair-memory retrieval as a unified adaptive policy that jointly considers coverage, quality and stage utility.

\subsection{Core Research Gaps in Retrieval Strategy}
Repair-memory retrieval essentially answers three core questions: \emph{which} memories to retrieve, \emph{where} to retrieve from, and \emph{when} to use them. Existing studies only address these questions separately, without forming an integrated adaptive framework. First, regarding memory selection, while semantic similarity-based code retrieval~\cite{husain2019codesearchnet} and in-context example selection~\cite{xiao2025diversity,li2025lail} have been widely studied, no system jointly evaluates candidates from relevance, utility, specificity and redundancy. Second, regarding retrieval scope, current systems mostly rely on repository-local histories; cross-project knowledge transfer verified in fault localization~\cite{chakraborty2025blaze,li2019deepfl} has not been used to dynamically adjust retrieval scope for memory-sparse repositories. Third, regarding stage matching, staged repair workflows imply phase-specific knowledge demands~\cite{li2025patchpilot,li2026conrad}, but existing stage-specific memory designs~\cite{shen2026structurally} are treated as independent storage schemes rather than an integrated part of the full retrieval pipeline. In summary, there lacks a unified framework that combines coverage-aware expansion, quality-aware selection and stage-aware routing into one coherent adaptive retrieval policy, which this work addresses with AdaRepair-Mem.
\section{Conclusion}
This paper studies memory augmented repository level program repair and identifies a central issue in existing systems. Historical repair experiences are useful, but they are often unevenly distributed across repositories, vary in quality, and do not match different repair stages well. To address this, we propose AdaRepair-Mem, denoted as AdaRepair-Mem, which integrates Coverage-aware Retrieval, Quality-aware Selection, and Stage-aware Routing to retrieve more suitable repair experiences for reproduction, localization, patch generation, patch refinement, and validation. Experiments on SWE-Bench-Lite and SWE-Bench-Verified show that AdaRepair-Mem improves repair performance, especially for repositories with limited historical experiences, and achieves better token efficiency than the base system.
\section{Data Availability}
\label{sec:data-availability}
An anonymous replication package is available for review.\footnote{\url{https://anonymous.4open.science/r/AdaRepair-Mem-EB46/}}
It contains the implementation, controlled benchmark, seeded and held-out
fault results, benign-transformation results, ablation data, diagnostic
and witness data, runtime data, plotting scripts, and reproduction
instructions. The package also includes the paper-facing CSV files used
to generate the figures, so the reported counts, ablations, diagnostics,
external-witness summaries, and runtime measurements can be checked
without rerunning the experiments.

\bibliographystyle{IEEEtran}
\bibliography{refs}
\clearpage

\end{document}